\documentclass[pre,twocolumn,aps,amsmath,showpacs,amsfonts,amssymb,10pt]{revtex4-1}
\usepackage{graphicx}
\usepackage{times}
\usepackage[utf8]{inputenc}
\usepackage[english]{babel}
\usepackage{fancyhdr}
\usepackage{marvosym}
\usepackage{hyperref}
\usepackage{amsmath}
\usepackage{mathrsfs}
\usepackage{latexsym}
\usepackage{amssymb}
\usepackage{graphicx} % Paquete que permite la incorporación de gráficas al documento
\usepackage[dvips,dvipsnames,usenames]{color} % Paquete que permite el uso de letras con colores
\usepackage{float} % Paquete que permite la orden [H] = Pon esta figura aquí bajo mi responsabilidad por muy feo que quede
\usepackage{bbm} 
\newcommand{\abs}[1]{\vert #1\vert}

\begin{document}

\title{Casimir repulsion between Topological Insulators in the diluted regime}

\author{Pablo Rodriguez-Lopez}
\affiliation{Departamento de F\'\i sica Aplicada I and GISC, Facultad de Ciencias F\'\i sicas, Universidad Complutense, 28040 Madrid, Spain}

\begin{abstract}
The Pairwise Summation Approximation (PSA) of Casimir energy is applied to a system of two dielectrics with magnetoelectric coupling. In particular, the case of Topological Insulators (TI) is studied in detail.
Depending on the the optical response of the TI, we obtain a stable equilibrium distance, atraction for all distances, or repulsion for all distances at zero temperature. This equilibrium distance disappears in the high temperature limit. These results are independent on the geometry of the TI, but are only valid in the diluted approximation.
\end{abstract}

%{42.50.Lc, 85.85.+j}

% 42.50.Lc   Quantum fluctuations, quantum noise, and quantum jumps
% 05.40.Ca   Noise: fluctuation phenomena,
% 03.70.+k   Quantized fields,
% 41.20.-q   Magnetic fields in electromagnetism,
% 05.20.-y   Statistical Mechanics
% 12.20.-m   Quantum electrodynamics (QED) in particle physics
% 45.70.-n   Granular systems, classical mechanics of
% 45.70.Mg   Mixing granular systems
% 85.85.+j   MEMS - Nanotechnology nanoelectromechanical systems, 
% 42.25.Fx   Optical diffraction
% 05.30.-d   Fermi-Dirac statistics,

\maketitle

\section{Introduction}
Since 1948, when Casimir introduced the energy that got his name \cite{Casimir Placas Paralelas}, calculation formulas have been looked for. Many analytical and numerical methods have been proposed, such as the zeta function technique, the heat kernel method, semiclassical methods or Green function (local) methods just to mention a few of them \cite{Review Casimir}. However, exact results have been obtained only for some simple geometries. In this context, several approximations have been proposed such as the Proximity Force Approximation or PSA, to cite two of them.

More recently, a novel Multiscattering formalism of Casimir effect for the electromagnetic (EM) field have been proposed~\cite{EGJK, Review-Jamal-Emig}. It had been successfully used in calculations of Casimir effect in new geometries, but has also been applied to obtain a generalization of PSA to dielectrics with magnetic response in \cite{PSA_EM_general}. In addition to that, this Multiscattering formalism was applied to obtain a generalization of Earnshaw's theorem to Casimir effect, claiming on the impossibility of equilibrium states in systems governed by Casimir physics~\cite{Casimir_No_Equilibrio_Estable}.

On the other hand, it has been suggested that TI could lead to the appearance of equilibrium systems because Casimir effect in \cite{PRL_Casimir_TI}.

Our goal is the generalization of PSA to obtain the Casimir energy between dielectrics with magnetoelectric response. To do so, we use the generalized constitutive relations of materials with magnetoelectric coupling to obtain a potential, which is the tree level of the Born series of the required $\mathbb{T}$ matrix.

We will apply this formalism to the Casimir energy between TI, as a result, we obtain a prelora of different behavior of the system depending on the electromagnetic response of the TI. Atraction for all distances, repulsion for all distances, or also the existence of a stable equilibrium distance are obtainable in this system at zero and finite temperature.

Then we find a system where the recent extension of Earnshaw's theorem of inexistence of stability on systems governed by Casimir physics~\cite{Casimir_No_Equilibrio_Estable} is not applicable (magnetoelectric coupling were not considered in~\cite{Casimir_No_Equilibrio_Estable}), therefore stable positions are not forbidden in the system studied here.

In addition to that, we also propose a system candidate to perform quantum levitation in vacuum. Other examples of repulsion within Casimir systems have been proposed for very specific geometries in vacuum~\cite{Casimir_Repulsion_metals_in_vacuum}, for systems immersed in fluids with dielectric properties different of vacuum~\cite{PhysRevLett.104.160402}, or for the Casimir interaction between dielectrics and metamaterials~\cite{Casimir_Metamateriales}.

In this article we extend the results for TI for arbitrary geometry in the PSA, valid only in the diluted limit. It is, when the dielectric constant $\epsilon_{\alpha}\to\epsilon_{0}$ and the diamagnetic constant $\mu_{\alpha}\to\mu_{0}$ for any $\alpha$ body of the system.

The article is structured as follows:
Using the generalized constitutive relations of materials with magnetoelectric coupling and the formalism presented in~\cite{PSA_EM_general}, in Sect. \ref{sec: 2} we will obtain the PSA of Casimir energy for the zero and high temperature limits starting from the exact Casimir energy formula given in \cite{EGJK}. In Sect. \ref{sec: 3} we will apply the formalism developed in Sect. \ref{sec: 2} to obtain the PSA Casimir energy for TI in the zero and high temperature limits. We will obtain conditions to the appearance of atraction, repulsion and the appearance of equilibrium distances in term of the response of the TI to the EM field. Finally, in Sect. \ref{sec: 4} we will perform a numerical study of the PSA Casimir energy between two TI parallel plates to compare the results of PSA with exact results already done in~\cite{PRL_Casimir_TI}, to study the validity of the approximations made. We also will obtain numerical results of the PSA Casimir energy for the sphere-plate system, because its experimental relevance~\cite{Casimir_Experimento_Cantilever}.

\section{Generalization of PSA to dielectrics with magnetoelectric coupling}\label{sec: 2}
Our goal is the calculation of the complete electromagnetic Casimir energy between two bodies in the soft dielectric limit with magnetoelectric response. For this purpose, we use the multiscattering approach to EM Casimir effect \cite{EGJK, Review-Jamal-Emig} to obtain the Casimir energy between two compact bodies at a given temperature $T$ as
\begin{equation}\label{Energy_finite_T}
E_{T} = k_{B}T{\sum_{n=0}^{\infty}}'\ln\abs{\mathbbm{1} - \mathbb{N}(\kappa_{n})}.
\end{equation}
where $\mathbbm{1}$ is the identity matrix, $\kappa_{n} = \frac{n}{\lambda_{T}}$ are the Matsubara frequencies and $\lambda_{T} = \frac{\hbar c}{2\pi k_{B}T}$ is the thermal wavelength. The prime indicates that the zero Matsubara frequency contribution has height of $1/2$. All the information regarding the system is described by the $\mathbb{N}$ matrix. For a system of two objects, this matrix is $\mathbb{N} = \mathbb{T}_{1}\mathbb{U}_{12}\mathbb{T}_{2}\mathbb{U}_{21}$. $\mathbb{T}_{i}$ is the T scattering matrix of the \textit{i}th object, which accounts for all the geometrical information and electromagnetic properties of the object. $\mathbb{U}_{ij}$ is the translation matrix of electromagnetic waves from object $i$ to object $j$, which accounts for all information regarding the relative positions between the objects of the system.
From Eq. \eqref{Energy_finite_T}, the quantum ($T\to 0$) and classical limits ($\hbar\to 0$, equivalent to the high temperature limit) are easily obtained as
\begin{equation}\label{Energy_zero_T}
E_{0} = \frac{\hbar c}{2\pi}\int_{0}^{\infty}dk\ln\abs{\mathbbm{1} - \mathbb{N}(\kappa)}
\end{equation}
and
\begin{equation}\label{Energy_high_T}
E_{cl} = \frac{k_{B}T}{2}\ln\abs{\mathbbm{1} - \mathbb{N}(0)}
\end{equation}
respectively. To obtain the PSA of Casimir energy, we apply the formalism developed in \cite{PSA_EM_general}. Then, using the relation $\ln\abs{A} =\textrm{Tr}(\ln(A))$ and that $\ln(1 - x) = - \sum_{p=1}^{\infty}\frac{x^{p}}{p}$, we write Eq. \eqref{Energy_finite_T} as the asymptotic series
\begin{equation}\label{Formula de Emig en forma de traza}
E_{T} = - k_{B}T{\sum_{n=0}^{\infty}}'\sum_{p=1}^{\infty}\frac{1}{p}\textrm{Tr}\left(\mathbb{N}^{p}(\kappa_{n})\right).
\end{equation}
In addition to that, we use the position representation of operators instead the multipole representation used in \cite{EGJK}, so we can identify $\mathbb{U}_{\alpha\beta} = G_{0\alpha\beta}$, where $G_{0\alpha\beta}$ is the matricial free dyadic Green function given in \cite{PSA_EM_general}.

The $\mathbb{T}$ operator is related with the potential $V$ by the Lippmann-Schwinger equation. When $V$ is small, we can apply a  Born expansion to the Lippmann-Schwinger equation to obtain an approximation for the $\mathbb{T}$ operator in the diluted limit as $\mathbb{T}_{i}\approx V_{i}$. Here we remind that the use of a Born approximation as a result of the Lippmann-Schwinger equation is more valid for lower potentials. In our case this means that PSA will be valid in the diluted limit, and we will need high order corrections when this diluted limit does not be longer applicable.

To define the potential $V_{i}$, we use the generalized constitutive relations of materials with magnetoelectric coupling
\begin{align}\label{generalized_constituve_relations}
\textbf{D} & = \epsilon\textbf{E} + \alpha\textbf{H},\nonumber\\
\textbf{B} & = \beta\textbf{E} + \mu\textbf{H},
\end{align}
where $\alpha$ and $\beta$ are the magnetoelectric couplings. The potential of each body can be defined as the difference of energy of the EM field because the existence of this body. Having into account that the energy of the EM field is defined as
\begin{equation}
E = \frac{1}{2} \int_{\Omega}dx^{\mu}\left(\textbf{E}\cdot\textbf{D} + \textbf{H}\cdot\textbf{B}\right),
\end{equation}
where $\textbf{D} = \epsilon_{0}\textbf{E}$ and $\textbf{H} = \mu_{0}\textbf{B}$ in the vacuum, the EM energy in presence of $N$ generalized dielectrics is
\begin{equation}
E = \frac{1}{2} \int_{\Omega} dx^{\mu}\left(\epsilon_{0}\textbf{E}^{2} + \mu_{0}\textbf{H}^{2}\right) + \sum_{i=1}^{N}\Delta E_{i}.
\end{equation}
We use the generalized constitutive relations given in Eq. \eqref{generalized_constituve_relations} to obtain the excess of energy because the existence of each dielectric as
\begin{align}
\Delta E_{i} & = \frac{1}{2}\int_{\Omega_{i}}dx^{\mu}\left(\textbf{E},\textbf{H}\right)\left(\begin{array}{c|c}
\left(\epsilon_{i} - \epsilon_{0}\right) & \alpha_{i}\\
\hline
\beta_{i} & \left(\mu_{i} - \mu_{0}\right)\end{array}\right)\left(\begin{array}{c}
\textbf{E}\\
\textbf{H}\end{array}\right),
\end{align}
then the potential is defined as
\begin{align}\label{Potencial_dielectric}
V_{i} & = \left(\begin{array}{c|c}
\tilde{\epsilon}_{i} & \alpha_{i}\\
\hline
\beta_{i} & \tilde{\mu}_{i}\end{array}\right)\chi_{i}\left(\textbf{r}\right),
\end{align}
where $\tilde{\epsilon}_{i} = \epsilon_{i} - \epsilon_{0}$, $\tilde{\mu}_{i} = \mu_{i} - \mu_{0}$ and $\chi_{i}\left(\textbf{r}\right)$ is the characteristic function of the $i$ body volume (1 inside the body and 0 in the rest of the space). The generalization to space dependent dielectric constants is straightforward.

Now we study the lowest expansion order in dielectric quantities of the Casimir energy. In the lowest expansion order ($p_{\text{Max}} = 1$ and $\mathbb{T}_{i} = V_{i}$ in Eq. \eqref{Formula de Emig en forma de traza}), we get the asymptotic approximation of the Casimir energy between two bodies as
\begin{equation}
E_{T} = - k_{B}T{\sum_{n=0}^{\infty}}'\textrm{Tr}\left(V_{1}G_{0,12}V_{2}G_{0,21}\right),
\end{equation}
where all operators depends on $\kappa_{n}$. These operators are defined over three different linear spaces \cite{PSA_EM_general}: (1) An $EH$ space, whose components are the electric and the magnetic field; (2) over the space coordinates, because we are working with a vector and a pseudovector; and (3) over positions. We must solve the trace over these three spaces: EH space, vector coordinate space, and position space and sum over the Matsubara frequencies to obtain the PSA energy.

As a first application, we use the potential $V_{i}$ given in Eq. \eqref{Potencial_dielectric} to obtain the PSA Casimir energy of frequency independent dielectric with magnetoelectric couplings at zero temperature as
\begin{equation}
E = \frac{-\hbar c}{(4\pi)^{3}}\gamma_{0}\int_{1}\int_{2}\frac{d\textbf{r}_{1}d\textbf{r}_{2}}{\abs{\textbf{r}_{1} - \textbf{r}_{2}}^{7}},
\end{equation}
where
\begin{align}
\gamma_{0} & = 23\tilde{\epsilon}_{1}\tilde{\epsilon}_{2} - 7\tilde{\epsilon}_{1}\tilde{\mu}_{2} - 7\tilde{\mu}_{1}\tilde{\epsilon}_{2} + 23\tilde{\mu}_{1}\tilde{\mu}_{2}\nonumber\\
 & + 7\alpha_{1}\alpha_{2} + 23\alpha_{1}\beta_{2} + 23\beta_{1}\alpha_{2} + 7\beta_{1}\beta_{2},
\end{align}
This result generalizes the Feinberg and Sucher potential~\cite{VdW int. magnetica} to objects with magnetoelectric couplings, but only in the diluted limit. In the high temperature limit, the PSA of Casimir energy is
\begin{equation}
E = \frac{- k_{B}T}{(4\pi)^{2}}\gamma_{cl}\int_{1}\int_{2}\frac{d\textbf{r}_{1}d\textbf{r}_{2}}{\abs{\textbf{r}_{1} - \textbf{r}_{2}}^{6}},
\end{equation}
where
\begin{equation}
\gamma_{cl} = 3\tilde{\epsilon}_{1}\tilde{\epsilon}_{2} + 3\tilde{\mu}_{1}\tilde{\mu}_{2} + 3\alpha_{1}\beta_{2} + 3\beta_{1}\alpha_{2}.
\end{equation}

It is a straightforward calculation obtain the PSA at any finite temperature. The result is similar to the finite temperature results of PSA for diluted dielectrics shown in~\cite{PSA_EM_general}.

As we can see, if magnetoelectric couplings have different signs and are strong enough to compensate the usual electric and magnetic coupling of dielectrics, a repulsive Casimir energy between objects can be achieved in the diluted limit.

\section{PSA for TI}\label{sec: 3}
Recently, in \cite{PRL_Casimir_TI}, A. Grushin and A. Cortijo demonstrated the existence of an equilibrium distance between topological insulators with opposite topological polarizabilities sign between parallel plates. This result was extended for all temperatures in \cite{PRB_Casimir_TI}, leading to a reduction of the equilibrium distance with the temperature until the disappearance of the equilibrium distance in the high temperature regime. Regions of Casimir repulsion for all distances were also reported for some values of $w = \frac{\omega_{e}}{\omega_{R}}$ and $\abs{\theta}$. In this part of the article, we apply the formalism developed in the previous section to TI. 
The constitutive relations of TI are $\textbf{D} = \epsilon\textbf{E} + \alpha\left(\theta/\pi\right)\textbf{B}$ and $\textbf{H} = \textbf{B}/\mu - \alpha\left(\theta/\pi\right)\textbf{E}$. In order to apply the formalism developed in the last section, we need the fields $\textbf{D}$ and $\textbf{B}$ as a function of $\textbf{E}$ and $\textbf{H}$, then
\begin{align}
\textbf{D} & = \left(\epsilon + \mu\alpha^{2}\left(\frac{\theta}{\pi}\right)^{2}\right)\textbf{E} + \mu\alpha\left(\frac{\theta}{\pi}\right)\textbf{H},\nonumber\\
\textbf{B} & = \mu\alpha\left(\frac{\theta}{\pi}\right)\textbf{E} + \mu\textbf{H},
\end{align}
where $\alpha$ is the fine structure constant ($\alpha = \frac{e^{2}}{\hbar c}$)
Following the discussion in \cite{PRB_Casimir_TI}, we assume $\mu = \mu_{0}$ and $\theta$ is quantized in odd integer values of $\pi$ such that
\begin{equation}
\theta = (2 n + 1)\pi,
\end{equation}
where $n\in\mathbb{Z}$, determined by the nature of the magnetic coating, but independent of the absolute value of the magnetization of the coating. Positive or negative values of $\theta$ are related to different signs of the magnetization on the surface \cite{Science_TI}, which we consider is perpendicular to the surface of the body. Being a topological contribution, $\theta$ is defined in the bulk as a constant whenever the bulk Brillouin zone is defined \cite{Brillouin}. %Surface effects are not taken into account in this model~\cite{Private_Communication}, but the model presented here can be easily generalized to cope with them. We just should add a surface contribution with the dependence on the imaginary frequency of the magnetoelectric coupling of the studied model, with a penetration length $\delta$ of the surface contribution. In the model studied here, $\delta = 0$ is assumed and therefore the surface contribution to Casimir energy is neglected.
In addition to that, we assume that the frequency dependent dielectric function $\epsilon(\omega)$ is described by an oscillator model of the form:
\begin{equation}
\epsilon(i\kappa) = \epsilon_{0} + \sum_{i}\frac{\omega_{e,i}^{2}}{\omega_{R,i}^{2} + \gamma_{R,i}c\kappa + c^{2}\kappa^{2}}.
\end{equation}
Because there are few experimental results of $\epsilon(i\kappa)$ for TI, we will assume that there is just one resonance. The case of multiple resonances or another model of $\epsilon(i\kappa)$ can be easily generalized~\cite{PRL_Casimir_TI}.

\begin{figure}[h]
\begin{center}
\includegraphics{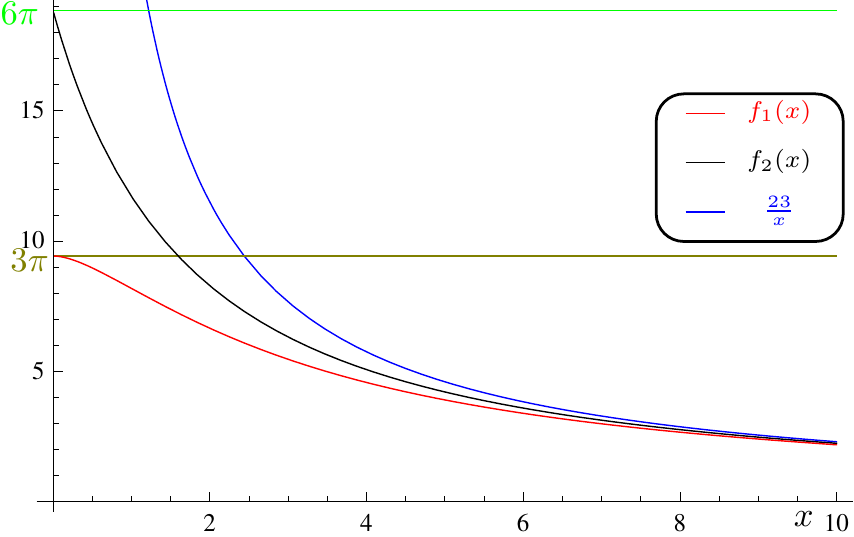}
\caption{\label{Plot_f00_f02} (Color online) $f_{1}(x)$ (red curve) and $f_{2}(x)$ (black curve) as a function of the dimensionless distance $x$. The blue curve is their common asymptotic limit to $\frac{23}{x}$ at large $x$. The limits at small $x$ of $f_{1}(x\to 0)\to 3\pi$ (yellow curve) and $f_{2}(x\to 0)\to 6\pi$ (green curve) are also plotted.}
\end{center}
\end{figure}

%\section{Asymptotic Casimir energy between 2 compact TI}\label{sec: 3.1}
Because the frequency dependence of $\epsilon(i\kappa)$ for TI, it is not possible to obtain the integrand of the PSA energy at zero temperature in a closed form in terms of simple functions. Then the PSA Casimir energy at zero temperature for topological insulators is given by the integral
\begin{equation}
E_{0}^{PSA} = - \frac{\hbar c}{(4\pi)^{3}}\int_{1}\int_{2}\frac{d\textbf{r}_{1}d\textbf{r}_{2}}{\abs{\textbf{r}_{1} - \textbf{r}_{2}}^{7}}\gamma_{0},
\end{equation}
where $\gamma_{0}$ now depends on the distance between points of the TI and on the dielectric properties of both TI as 
\begin{align}\label{Integrando_PSAE_for_TI}
\gamma_{0} & = 
w_{1}^{2}w_{2}^{2} x f_{1}(x) + 60 \bar{\alpha}_{1}\bar{\alpha}_{2} + 23\bar{\alpha}_{1}^{2} \bar{\alpha}_{2}^{2}\nonumber\\
&  + \left(\bar{\alpha}_{1}^{2}w_{2}^{2} + \bar{\alpha}_{2}^{2}w_{1}^{2}\right)x f_{2}(x),
\end{align}
where $\bar{\alpha}_{i} = \alpha\frac{\theta_{i}}{\pi}$, $w = \frac{\omega_{e}}{\omega_{R}}$, $x = \frac{\omega_{R}}{c}\abs{\textbf{r}_{1} - \textbf{r}_{2}}$ is the dimensionless distance and $f_{1}(x)$ and $f_{2}(x)$ are functions which depends on a linear combination of different Meijer $G$ functions too long to be written here. All terms of Eq.~\eqref{Integrando_PSAE_for_TI} are positive ($f_{1}(x)$ and $f_{2}(x)$ are smooth positive functions in the real positive axis, see Fig.~\ref{Plot_f00_f02}) for any value of $\theta_{1}$ and $\theta_{2}$, except the product $\bar{\alpha}_{1}\bar{\alpha}_{2}$, which becomes negative when $\text{sign}(\theta_{1}) = - \text{sign}(\theta_{2})$.

At short distances we can approximate $f_{1}(x)\approx 3\pi$ and $f_{2}(x)\approx 6\pi$, then $\gamma_{0}$ can be approximated by
\begin{equation}\label{Integrando_PSAE_for_TI_short}
\gamma_{0} = 60\bar{\alpha}_{1}\bar{\alpha}_{2} + 23 \bar{\alpha}_{1}^{2}\bar{\alpha}_{2}^{2}.
\end{equation}
If $\text{sign}(\theta_{1}) = - \text{sign}(\theta_{2})$, the energy $E_{0}^{PSA}$ becomes positive (i.e. repulsive) for all
\begin{equation}\label{Repulsion_condition_short_distance_limit}
\theta_{1}\theta_{2} < - \frac{60\pi^{2}}{23\alpha^{2}}.
\end{equation}
For diluted materials, this condition is beyond the positive energy condition $\abs{\theta_{1}\theta_{2}} < \frac{\pi^{2}}{\alpha^{2}}\sqrt{\epsilon_{1}(0)\epsilon_{2}(0)} < \frac{60\pi^{2}}{23\alpha^{2}}$ \cite{Positive_Energy_Condition}, then we can consider it is always fulfilled for diluted dielectrics.

At large distances we can approximate $f_{1}(x)\approx\frac{23}{x}$ and $f_{2}(x)\approx\frac{23}{x}$, then $\gamma_{0}$ can be approximated by
\begin{equation}\label{Integrando_PSAE_for_TI_large}
\gamma_{0} = w_{1}^{2}w_{2}^{2} + \bar{\alpha}_{1}^{2}\bar{\alpha}_{2}^{2} + \bar{\alpha}_{1}^{2}w_{2}^{2} + \bar{\alpha}_{2}^{2}w_{1}^{2} + \frac{60}{23}\bar{\alpha}_{1}\bar{\alpha}_{2}.
\end{equation}
If $\text{sign}(\theta_{1}) = - \text{sign}(\theta_{2})$ and $w_{1} = w_{2} = w$, the energy becomes positive (i.e. repulsive) for all
\begin{equation}\label{Repulsion_condition_large_distance_limit}
w^{2} < \frac{23}{46}\left(- \bar{\alpha}_{1}^{2} - \bar{\alpha}_{2}^{2} + \sqrt{\bar{\alpha}_{1}^{4} + \bar{\alpha}_{2}^{4} - \frac{240}{23}\bar{\alpha}_{1}\bar{\alpha}_{2} - \frac{46}{23}\bar{\alpha}_{1}^{2}\bar{\alpha}_{2}^{2}}\right).
\end{equation}

As a consequence, there are several different regimes for the system in the quantum limit. When $\text{sign}(\theta_{1}) = \text{sign}(\theta_{2})$, the Casimir energy is enlarged because the contribution of topological polarizability $\theta$, but when $\text{sign}(\theta_{1}) = - \text{sign}(\theta_{2})$, different regimes appear.

When $\text{sign}(\theta_{1}) = - \text{sign}(\theta_{2})$ and for low enough absolute values of $\theta_{1}$ and $\theta_{2}$, the condition given in Eq. \eqref{Repulsion_condition_short_distance_limit} is fulfilled while the condition given in Eq. \eqref{Repulsion_condition_large_distance_limit} is not, then we have a repulsive Casimir energy at short distances and an attractive Casimir energy at large distances, then there must exist a stable equilibrium distance in this case.

But if  the condition given in Eq. \eqref{Repulsion_condition_large_distance_limit} is also fulfilled, the magnitude of the positive arguments of Eq. \eqref{Integrando_PSAE_for_TI} is not large enough to compensate the repulsion because topological charges $\theta$, then we obtain repulsion for all distances.

\begin{figure}[h]
\begin{center}
\includegraphics{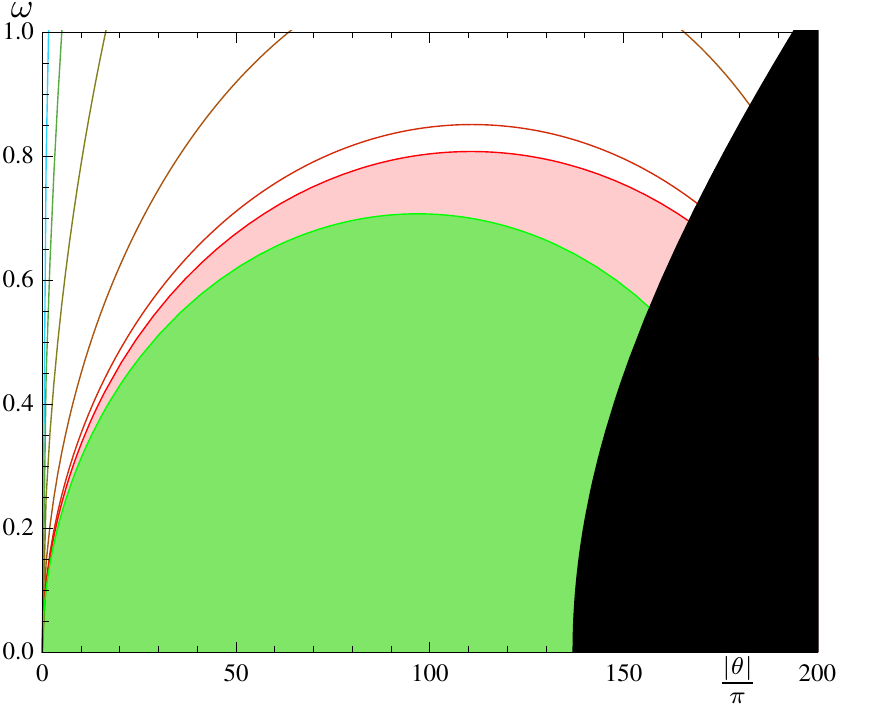}
\caption{\label{Mapa_Fases_TI}(Color online) Atraction versus Repulsion in the classical limit ($T\to\infty$) and in the quantum limit ($T\to 0$) as a function of $w$ and $\abs{\theta}$. For the Classical limit, the green region shows repulsion for all distances and the rest of phase space shows atraction for all distances. For the quantum limit, the red region shows repulsion for all distances and the white region shows parameters for which an equilibrium distance appears. Several curves of constant equilibrium dimensionless distance $x_{eq}$ are also plotted. The black region is a forbidden region for the parameters due to the positive energy condition $\abs{\bar{\alpha}} < \sqrt{\epsilon(0)}$ \cite{Positive_Energy_Condition}.}
\end{center}

\end{figure}

In the high temperature limit, another result is obtained. In this case the PSA gives the Casimir energy as
\begin{equation}
E_{cl}^{PSA} = - \frac{3 k_{B}T}{(4\pi)^{2}}\gamma_{cl}\int_{1}\int_{2}\frac{d\textbf{r}_{1}d\textbf{r}_{2}}{\abs{\textbf{r}_{1} - \textbf{r}_{2}}^{6}},
\end{equation}
where, contrary to the zero temperature case, $\gamma_{cl}$ does not depend on the distance between points, and it is given by
\begin{equation}
\gamma_{cl} = w_{1}^{2}w_{2}^{2} + \bar{\alpha}_{1}^{2}w_{2}^{2} + \bar{\alpha}_{2}^{2}w_{1}^{2} + \bar{\alpha}_{1}^{2}\bar{\alpha}_{2}^{2} + 2\bar{\alpha}_{1}\bar{\alpha}_{2}.
\end{equation}
Depending on the values of $w$ and $\abs{\theta}$, we obtain or repulsion for all distances or atraction for all distances. The condition to be fulfilled in order to obtain repulsion in the classical limit, when $w_{1} = w_{2} = w$ is
\begin{equation}\label{Repulsion_condition_classical_limit}
w^{2} < \frac{1}{2}\left(- \bar{\alpha}_{1}^{2} - \bar{\alpha}_{2}^{2} + \sqrt{\bar{\alpha}_{1}^{4} + \bar{\alpha}_{2}^{4} - 8\bar{\alpha}_{1}\bar{\alpha}_{2} - 2\bar{\alpha}_{1}^{2}\bar{\alpha}_{2}^{2}}\right).
\end{equation}
In Fig. \ref{Mapa_Fases_TI}, we represent the behavior of the PSA energy as a function of $\theta$ and $w$ when $w_{1} = w_{2}$ and $\theta_{1} = \theta = - \theta_{2}$ in the quantum and in the classical limit, having into account there exist a forbidden region because the positive energy condition \cite{Positive_Energy_Condition}. In both cases we find different regions of parameters of repulsion for all distances, but in the quantum limit there is a region of existence of an equilibrium distance, while in the classical limit a region of atraction for all distances appears.

It is expected that, at finite temperatures, the equilibrium distances found in the quantum limit would be reduced when temperature increases~\cite{PRB_Casimir_TI}, and Fig. \ref{Mapa_Fases_TI} suggest that behavior, but we have let this study as a future work.

\section{Casimir energy between spheres and plates}\label{sec: 4}
%A claim would be done to the method performed here. It is a method just valid in the diluted limit. Then the results presented here are general in the geometry of the bodies, but not in the dielectric response of materials.

In this section we will use the PSA formulas presented in the previous one to obtain the Casimir energy to specific systems. In particular, we study the sphere-plate system because it experimental relevance and the two infinite plates system to compare the PSA results with the exact ones. Both systems were studied at zero temperature.

We were not able to obtain analytical results, so a numerical integration procedure has been implemented.

\begin{figure}[h]
\begin{center}
\includegraphics{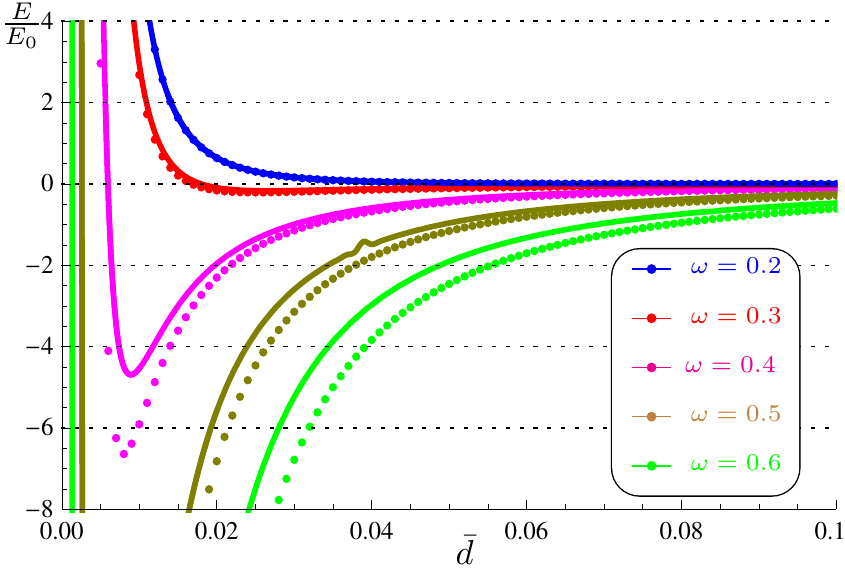}
\caption{\label{Comparacion_PSA_Multiscattering}(Color online) Comparison between exact Casimir energies in units of $E_{0} = A\hbar c/(2\pi)^{2}\left(\omega_{R}/c\right)^{3}$(full curves) and PSA (points) at zero temperature for TI infinity plates with $\theta_{1} = - \theta_{2} = \pi$ and different $\epsilon(0)$ as a function of the dimensionless distance $\bar{d}$~\cite{PRL_Casimir_TI}. PSA tends to overestimate the absolute value of the energy, but captures the equilibrium distance without appreciable error. The overestimation of the energy increases with $\epsilon(0)$, as expected because the nature of the approximation.}
\end{center}
\end{figure}

The two infinite parallel TI plates has been already studied in \cite{PRL_Casimir_TI}. Our interest here is to compare the results obtained by the PSA with the exact ones. We impose $\theta_{1} = - \theta_{2} = \pi$ in order to obtain an equilibrium distance, and different $w$ from $w = 0.2$ to $w = 0.6$. The obtained results of PSA and the exact ones are compared in Fig. \ref{Comparacion_PSA_Multiscattering}, where we find that PSA tends to overestimate the magnitude of the Casimir energy the more the larger $w$. This fact reflects the nature of the approximation made in PSA, where we must assume the diluted limit is valid. On the other side, an excellent approximation of the equilibrium distance between plates is obtained, then the PSA gives a good qualitative result.

\begin{figure}[H]
\begin{center}
\includegraphics{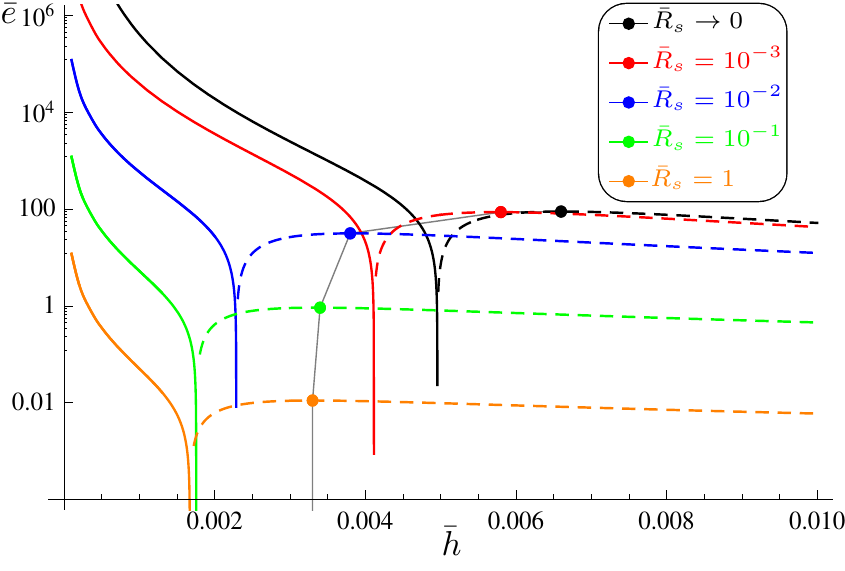}
\caption{\label{Figura_Esfera_Placa}(Color online) Dimensionless PSA Casimir energies in the sphere-plate TI system per unit of volume of the sphere $\bar{e} = \frac{c^{3}}{\hbar\omega_{R}^{4}}\frac{E}{V}$ with $\theta_{1} = - \theta_{2} = \pi$ and $w = 0.45$ as a function of the dimensionless distance between the plate and the contact point of the sphere $\bar{h} = \bar{d} - \bar{R}_{s}$, where $\bar{d}$ is the dimensionless distance from the surface of the plate to the center of the sphere. Positive energies are the whole curves, while negative energies are the dashed curves. The equilibrium distance for each sphere radius is represented by a dot. These dots are joined by a soft line to have an idea of the dependence of the equilibrium distance with the radius of the sphere.}
\end{center}
\end{figure}

We have also used the PSA to study the sphere-plate system because it experimental relevance in Casimir effect experiments. In this case, we also impose $\theta_{1} = - \theta_{2} = \pi$ and $w = 0.45$ for both plate and sphere in order to obtain an equilibrium distance at $T=0$ (see Fig. \ref{Mapa_Fases_TI}). We vary the dimensionless radius of the sphere $\bar{R}_{s} = \frac{\omega_{R}}{c}R_{s}$ from $\bar{R}_{s}\to 0$ until $\bar{R}_{s} = 1$ to observe it effect in the equilibrium distance. In Fig.~\ref{Figura_Esfera_Placa}, the PSA energy per unit of volume of the sphere is plotted as a function of the distance between nearest points of the plate and the sphere. As a result, an equilibrium distance has been obtained for all studied sphere radii. This equilibrium distance reduces when the radius increases until reaching a constant value. The appearance of an equilibrium distance is easy to understand because the nature of the integrand in PSA calculations at zero temperature. As discussed in Eq. \eqref{Integrando_PSAE_for_TI}, from a given distance between sphere and plate, the points of the sphere nearer to the plate tends to increase the Casimir energy (giving a repulsive contribution), while the rest of points of the sphere tends to reduce this energy (giving an atractive contribution). As at short distances Eq. \eqref{Integrando_PSAE_for_TI} diverges as $\frac{1}{d^{7}}$ and $\int_{0}^{d}\frac{dx}{x^{7}}$ diverges, it is evident that there always be a distance where repulsion compensates atraction, reaching the system an equilibrium distance.

\section{Conclusions}\label{sec: 5}
In this article we have extended the PSA to dielectrics with magnetoelectric coupling. The extension to anisotropic materials of this formalism is straightforward, although we have not done it here, is shown how to do it for the case where it would be needed.

There exist several examples of magnetoelectric materials, as metamaterials~\cite{Casimir_Metamateriales}, already proposed as candidates of repulsive Casimir effect, and TI~\cite{PRL_Casimir_TI}, which have been proposed recently to obtain equilibrium systems in Casimir physics.

We remind here that the extension of Earnshaw's theorem to Casimir effect~\cite{Casimir_No_Equilibrio_Estable} does not forbid the appareance of equilibrium distances between TI, because magnetoelectric couplings were not considered.

As a result, we have demonstrated that two TI or arbitrary shape with opposed signs $\text{sign}(\theta_{1}) = - \text{sign}(\theta_{2})$ can present or repulsion for all distances or the appearance of a stable equilibrium distance at zero temperature as a function of their dielectric properties, given by parameters $w$ y $\abs{\theta}$ in the model studied here.

At high temperatures the behavior of the system differs. Equilibrium distance collapses to zero, and the repulsion parameters range is reduced. Then we obtain, as a function of the dielectric properties of the TI, or atraction or repulsion for all distances.

We also have used the results derived here to study the Casimir energy for a system of two infinity TI plates, to compare the PSA with exact results given in \cite{PRL_Casimir_TI}. As a result, we found that PSA gives better quantitative results for lower $w$, as expected because PSA works better in the diluted limit. On the other side, we obtain an excellent approximation of the equilibrium distance between plates, then we obtain good qualitative results for all $\theta$ and $w$ studied.

We have also used the PSA to study the Casimir energy between an infinite TI plate and a TI sphere because the experimental relevance of this geometry in experiments of Casimir effect~\cite{Casimir_Experimento_Cantilever}.

The results presented in this article are valid for objects of arbitrary shape, but in the diluted limit. Even so, this is the relevant limit because topological polarizabilities $\theta$ scale with the fine structure constant $\alpha$, and Casimir repulsion and Casimir equilibrium are more relevant when $\epsilon$ is small too, in the diluted limit itself.

\acknowledgments
P. R.-L. acknowledges helpful discussions with R.~Brito, A. G. Grushin and A. Cortijo. This research was supported by the projects MOSAICO and MODELICO and a FPU MEC grant.

%\newpage
%\nocite*
%Para Bibtex, se usan las siguientes instrucciones:
%\bibliography{References}
%\bibliographystyle{unsrt}
% Para el caso que nos ocupa, lo haré de la forma cutre:

\end{document}